\documentclass[%
reprint,
superscriptaddress,
 amsmath,amssymb,
 aps,
 onecolumn, prfluids,
]{revtex4}

\usepackage[utf8]{inputenc}
\usepackage{amsthm}
\usepackage{bbm}
\usepackage{graphicx}
\usepackage{color}
\usepackage{upgreek}
\usepackage{bm}
\usepackage{textcomp}
\usepackage[usenames,dvipsnames]{xcolor}
\usepackage[normalem]{ulem}
\usepackage{wasysym}
\usepackage{siunitx}
\usepackage[linkcolor = blue, citecolor = blue, urlcolor = blue, colorlinks = true]{hyperref}

\begin{document}

\title{Flow rate--pressure drop relation for deformable channels \\ via fluidic and elastic reciprocal theorems}

\author{Evgeniy Boyko}
\email{eboyko@princeton.edu, eboyko@purdue.edu}
\affiliation{Department of Mechanical and Aerospace Engineering, Princeton University, New Jersey 08544, USA}
\affiliation{School of Mechanical Engineering, Purdue University, West Lafayette, Indiana 47907, USA}
\affiliation{Davidson School of Chemical Engineering, Purdue University, West Lafayette, Indiana 47907, USA}
\author{Howard A. Stone}
\email{hastone@princeton.edu}
\affiliation{Department of Mechanical and Aerospace Engineering, Princeton University, New Jersey 08544, USA}
\author{Ivan C. Christov}
\email{christov@purdue.edu}
\affiliation{School of Mechanical Engineering, Purdue University, West Lafayette, Indiana 47907, USA}

\date{\today}

\begin{abstract}

Viscous flows through configurations manufactured or naturally assembled from soft materials apply both pressure and shear stress at the solid–liquid interface, leading to deformation of the fluidic conduit's cross-section, which in turn affects the flow rate–pressure drop relation. Conventionally, calculating this flow rate–pressure drop relation requires solving the complete elastohydrodynamic problem, which couples the fluid flow and elastic deformation. In this Letter, we use the reciprocal theorems for Stokes flow and linear elasticity to derive a closed-form expression for the flow rate–pressure drop relation in deformable microchannels, bypassing the detailed calculation of the solution to the fluid-structure-interaction problem. For small deformations (under a domain perturbation scheme), our theory provides the leading-order effect of the interplay between the fluid stresses and the compliance of the channel on the flow rate–pressure drop relation. Our approach uses solely the fluid flow solution and the elastic deformation due to the corresponding fluid stress distribution in an undeformed channel, eliminating the need to solve the coupled elastohydrodynamic problem. Unlike previous theoretical studies that neglected the presence of lateral sidewalls (and considered shallow geometries of effectively infinite width), our approach allows us to determine the influence of confining sidewalls on the flow rate–pressure drop relation. In particular, for the flow-rate-controlled situation and the Kirchhoff--Love plate-bending theory for the elastic deformation, we show a trade-off between the effect of compliance of the deforming top wall and the drag due to sidewalls on the pressure drop. While compliance decreases the pressure drop, the drag due to sidewalls increases it. Our theoretical framework may provide insight into existing experimental data and pave the way for the design of novel optimized soft microfluidic configurations of different cross-sectional shapes.
\end{abstract}

\maketitle

\section{Introduction}
Pressure-driven viscous flows through conduits manufactured from soft materials apply both pressure and shear stress at the solid–liquid interface, leading to deformation of the cross-section, which in turn affects the relationship between the flow
rate $q$ and the pressure drop $\Delta p$ \cite{gervais2006flow,seker2009nonlinear,christov2021soft}. Understanding this $q-\Delta p$ relation is important for various microfluidic, lab-on-a-chip, biomedical, and soft robotics applications, such as pressure-actuated valves \cite{TMQ02}, passive fuses \cite{GMV17}, pressure sensors \cite{HHM02,OYE13}, soft actuators \cite{Polygerinos17,MEG17}, and estimating the drug injection force \cite{vurgaft2019forced}.
Conventionally, calculating the flow rate–pressure drop relation requires solving the complete elastohydrodynamic problem \cite{chakraborty2012fluid}, which couples the hydrodynamics to the elastic response, or a pressure-deformation relation is assumed \textit{a priori} \cite{rubinow1972flow,shapiro1977steady}. For example, recent studies used lubrication theory and linear elasticity to obtain the solution of the coupled problem for the fluid velocity and the elastic deformation. Then, the $q-\Delta p$ relation for Newtonian and complex fluids was derived from the latter for deformable microchannels that are slender and shallow \cite{christov2018flow,shidhore2018static,anand2019non,mehboudi2019experimental,wang2019theory,ramos2021fluid}. However, as we show, these detailed calculations of the solution for coupled fluid–structure interaction can be bypassed, at least in some cases, by jointly applying the reciprocal theorems for the fluidic and elastic problems.

The Lorentz reciprocal theorem has been applied widely in low-Reynolds-number fluid mechanics to facilitate the calculation of integrated quantities by eliminating the need for calculating the detailed velocity and pressure fields (e.g., \cite{lorentz1896general,happel1983,masoud2019reciprocal}). In particular, several studies showed the versatility of the Lorentz reciprocal theorem for evaluating the force and torque on a particle moving in the vicinity of a deformable boundary \cite{berdan1982motion,shaik2017motion,rallabandi2017rotation,rallabandi2018membrane,daddi2018reciprocal,kargar2021lift,bertin2022soft}, as well as its linear and angular velocities. Although the integral form of the reciprocal theorem is particularly convenient for calculating integrated hydrodynamic quantities such as force, torque, and flow rate~\cite{masoud2019reciprocal}, its use has been primarily limited to obtaining the force and torque acting on particles in flows of viscous fluids in unbounded and semi-infinite domains~\cite{leal1979motion,leal1980particle}. To date, only a few studies have utilized the reciprocal theorem to obtain the flow rate or flow rate--pressure drop relation for confined viscous Newtonian and complex fluid flows, such as in rigid channels~\cite{day2000lubrication,michelin2015reciprocal,boyko2021RT,boyko2022pressure}.

In addition to fluid mechanics, the reciprocal theorem has been used extensively in the solid mechanics community since \citet{maxwell1864calculation} and \citet{betti1872teoria}; we refer the interested reader to the book of \citet{achenbach2003reciprocity}. Similar to the fluidic reciprocal theorem, which relates the velocity $\boldsymbol{v}$ and stress  $\boldsymbol{\sigma}_{f}$ fields of one problem to the velocity $\hat{\boldsymbol{v}}$ and stress  $\hat{\boldsymbol{\sigma}}_{f}$ fields of an auxiliary problem, the elastic reciprocal theorem relates the displacement $\boldsymbol{u}$ and stress  $\boldsymbol{\sigma}_{s}$ fields of one problem to the displacement $\hat{\boldsymbol{u}}$ and stress $\hat{\boldsymbol{\sigma}}_{s}$  fields of an auxiliary problem. 
Given this similarity between the fluidic and elastic reciprocal theorems, one would expect to find the application of the theorems to fluid–structure interaction problems involving fluid flow and elastic deformation. However, to the best of our knowledge, no simultaneous application of the fluid and elastic reciprocal theorems has been presented to date, particularly, for calculating the flow rate–pressure drop relation for deformable channels.

In this Letter, we show how the reciprocal theorems for Stokes flow and linear elasticity can be harnessed to obtain the flow rate--pressure drop relation in deformable channels of initially rectangular cross-section, bypassing the detailed calculation of the solution to the fluid-structure interaction problem. Employing the slenderness of the geometry and considering small deformations, we derive a closed-form expression for the flow rate--pressure drop relation under a domain perturbation scheme. This relation accounts for the leading-order effect of the interplay between the fluid stresses and the compliance of the channel. Our approach uses only the fluid flow solution and the elastic deformation due to the corresponding fluid stress distribution in an undeformed channel, without the need to solve the coupled elastohydrodynamic problem. Furthermore, we show that our theory allows determining the influence of confining lateral sidewalls on the $q-\Delta p$ relation, in contrast to previous theoretical studies that neglected the presence of sidewalls and considered shallow geometries of effectively infinite width \cite{christov2018flow,shidhore2018static,mehboudi2019experimental,wang2019theory,martinez2020start,ramos2021fluid}. We illustrate the use of our approach for the model case of a thin deformable top wall that obeys the Kirchhoff--Love plate-bending theory. For the flow-rate-controlled situation, we show that, while increased compliance of the channel decreases the pressure drop, the drag due to the sidewalls increases it.

\section{Problem formulation and governing equations}

\begin{figure}[t]
 \centerline{\includegraphics[scale=1.1]{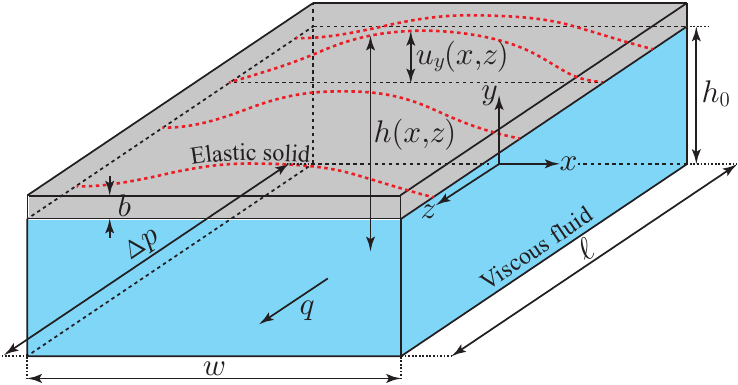}}\caption{Schematic illustration of the geometry consisting of a three-dimensional deformable channel of length $\ell$ with an initially rectangular cross-section of width $w$ and height $h_{0}$. The channel contains a viscous fluid steadily driven by an imposed flow rate $q$, leading to the deformation $u_{y}(x,z)$ of the top wall of thickness $b$, which in turn affects the pressure drop $\Delta p$ over the axial distance $\ell$. The sidewalls are assumed to be rigid.}
\label{F1}
\end{figure}

Consider the incompressible steady pressure-driven flow of a Newtonian viscous fluid in a slender and deformable channel of length $\ell$, width $w$, and (deformed) height $h$, as shown in Fig.~\ref{F1}. The fluid flow has velocity $\boldsymbol{v}=(v_{x},v_{y},v_{z})$ and pressure distribution $p$, which are induced by the imposed flow rate $q$. Our goal is to determine the resulting axial pressure drop $\Delta p$ for a given $q$. The top wall of the channel is soft and may deform due to the fluid stress distribution, acting at the fluid--solid interface, leading to the displacement field $\boldsymbol{u}=(u_{x},u_{y},u_{z})$. The sidewalls are assumed to be rigid. Specifically, we use $u_{y}^0(x,z)$ to denote the steady vertical displacement of the \emph{lower surface} of the top wall, i.e., the fluid--solid interface, so that its position is given by $y=h(x,z)=h_{0}+u_{y}^0(x,z)$, where $h_{0}$ is the undeformed height of the channel (in the absence of flow). We further assume the top wall of the channel has a constant thickness $b$ and constant material properties.

We consider low-Reynolds-number flow, so that fluid inertia is negligible compared to viscous forces. In this limit, the continuity and momentum equations governing the fluid motion take the form
\begin{equation}
\boldsymbol{\nabla\cdot v}=0,\qquad\boldsymbol{\nabla\cdot}\boldsymbol{\sigma}_{f}=\boldsymbol{0},
\label{Continuity+Momentum}
\end{equation}
where $\boldsymbol{\sigma}_{f}=-p\boldsymbol{I}+\mu(\boldsymbol{\nabla}\boldsymbol{v}+(\boldsymbol{\nabla}\boldsymbol{v})^{\mathrm{T}})$ is the Newtonian stress tensor, $\boldsymbol{I}$ is the identity tensor, and $\mu$ is the fluid's dynamic viscosity. The governing equations (\ref{Continuity+Momentum}) are supplemented by the no-slip and no-penetration boundary conditions along the channel walls, $\boldsymbol{v}=\boldsymbol{0}$ at $y=0,\,h(x,z)$, and at $x=\pm w/2$. Further, the integral constraint $\int_{-w/2}^{w/2}\int_{0}^{h(x)}v_z(x,y,z)\,\mathrm{d}y\,\mathrm{d}x=q$ enforces the flow rate. In addition, we assume that fluid exits at the outlet to atmosphere and set $p(\ell)=0$, so that $\Delta p=p(0)$.

Suppose that the top wall of the channel can be modeled as a linearly elastic isotropic solid with shear modulus $G$ and first Lam\'e parameter $\lambda$. Neglecting body forces in the solid, the steady stress balance in the elastic material takes the form
\begin{equation}
\boldsymbol{\nabla\cdot\sigma}_{s}=\boldsymbol{0},\label{Elalstic stress balance}
\end{equation}
where $\boldsymbol{\sigma}_{s}=\lambda(\boldsymbol{\nabla\cdot}\boldsymbol{u})\boldsymbol{I}+2G\boldsymbol{D}$ is the stress tensor of a linearly elastic solid and $\boldsymbol{D}=(\boldsymbol{\nabla}\boldsymbol{u}+(\boldsymbol{\nabla}\boldsymbol{u})^{\mathrm{T}})/2$ is the infinitesimal strain tensor. Note that $G$ and $\lambda$ are related to the Young's modulus $E$ and Poisson's ratio $\nu$ as $G=E/[2(1+\nu)]$ and $\lambda=E\nu/(1+\nu)(1-2\nu)$. The governing equation (\ref{Elalstic stress balance}) is supplemented by the no-displacement boundary condition, $\boldsymbol{u}=\boldsymbol{0}$ along the lines of contact, $z=0,\,\ell$, and $x=\pm w/2$, with the inlet/outlet and rigid walls. In addition, at the fluid--solid interface $y=h(x,z)$, the continuity of stresses requires that $\boldsymbol{n}\boldsymbol{\cdot}\boldsymbol{\sigma}_{s}=\boldsymbol{n}\boldsymbol{\cdot}\boldsymbol{\sigma}_{f}$, where $\boldsymbol{n}$ is the unit normal vector to the fluid--solid interface. This condition couples the fluidic and elastic problems. Finally, we assume a stress-free condition at the upper surface of the top wall, i.e., $\tilde{\boldsymbol{n}}\boldsymbol{\cdot}\boldsymbol{\sigma}_{s}=\boldsymbol{0}$, where $\tilde{\boldsymbol{n}}$ is the unit normal vector to the upper surface of the soft wall.

We introduce dimensionless variables based on lubrication theory,\begin{equation}
X=\frac{x}{w},\quad Y=\frac{y}{h_{0}},\quad Z=\frac{z}{\ell},\quad V_{x}=\frac{v_{x}}{(\epsilon/\delta)v_{c}},\quad V_{y}=\frac{v_{y}}{\epsilon v_{c}},\quad V_{z}=\frac{v_{z}}{v_{c}}, \quad P=\frac{p}{p_{c}},\quad H=\frac{h}{h_{0}},\quad U=\frac{u_{y}}{u_{c}},
\label{Non-dimensional variables}
\end{equation}
where $v_{c}=q/(h_{0}w)$ is the characteristic axial velocity scale, $p_{c}=\mu v_{c}\ell/h_{0}^{2}=\mu q\ell/(wh_{0}^{3})$ is the characteristic pressure scale, and $u_{c}$ is the characteristic scale of deformation of the top wall \cite{christov2018flow}. Also, we have defined $\epsilon=h_{0}/\ell$ and $\delta=h_{0}/w$, which represent, respectively, the slenderness and the shallowness of the channel. We assume $\epsilon=h_{0}/\ell$ to be small, $\epsilon\ll1$, but $\delta=h_{0}/w$ can be $\mathrm{O}(1)$, with the ordering $0<\epsilon\ll\mathrm{min}(\delta,1)$. Thus, unlike previous theoretical studies that assumed $\delta=h_{0}/w\ll1$ \cite{christov2018flow,shidhore2018static,mehboudi2019experimental,wang2019theory,martinez2020start,ramos2021fluid}, we consider a slender channel that is \emph{not} necessarily shallow.

In recent studies it was shown that for rectangular elastic top wall geometries, the  horizontal displacements $u_{x}$ and $u_{z}$ are much smaller in comparison to the vertical displacement $u_{y}$ \cite{wang2019theory}. The latter can also be rationalized using a scaling argument. Balancing the elastic elongational stress and the viscous stress at the top wall, we obtain that $u_{z}$  scales as $\mu q\ell/(Ewh_{0}^{2})$. Thus, for example, for the plate-bending theory in which $b\ll w$ and $u_{y}\sim\mu q\ell w^{3}/(Eb^{3}h_{0}^{3})$, we find $u_{z}/u_{y}\sim b^{3}h_{0}/w^{4}\ll1$. A similar argument yields $u_{x}/u_{y}\ll1$.
Therefore, it is sufficient and convenient, to consider that the entire deformation of the top wall is in the $y$-direction. Further, for thin structures $b\lesssim w$, the deformation at the fluid--solid interface is representative of the entire wall motion (i.e., $u_y\approx u_y^0$), consistent with reduced theories of elastic deformation, such as those due to Winkler, Kirchhoff--Love, Mindlin--Reissner, and F\"oppl--von K\'arm\'an \cite{timoshenkoPlates,howell2009applied}. Therefore, in the following, we make explicit the kinematic assumption that the displacement of the fluid--solid interface can be written as $\boldsymbol{u}=(0,u_{y}(x,z),0)$. Thus, the dimensionless deformed shape of the channel $H(X,Z)$ can be expressed in terms of the dimensionless top wall deformation $U(X,Z)$ as 
\begin{equation}
H(X,Z)=1+(u_{c}/h_{0})U(X,Z)=1+\beta U(X,Z),\label{Non-dimensional shape H}
\end{equation}
where $\beta=u_{c}/h_{0}$ is the dimensionless ratio that
quantifies the compliance of the top wall.

\section{Reciprocal theorems for viscous flows in weakly deformable channels}

\subsection{Fluidic reciprocal theorem}

Let $\hat{\boldsymbol{v}}$, $\hat{p}$, and $\hat{\boldsymbol{\sigma}}_{f}$
denote, respectively, the velocity, pressure, and stress fields corresponding
to the solution of the pressure-driven flow in a rigid (rectangular)
channel, satisfying $\boldsymbol{\nabla\cdot}\hat{\boldsymbol{v}}=0$, $\boldsymbol{\nabla\cdot}\hat{\boldsymbol{\sigma}}_{f}=\boldsymbol{0}$, with $\hat{\boldsymbol{\sigma}}_{f}=-\hat{p}\boldsymbol{I}+\mu(\boldsymbol{\nabla}\hat{\boldsymbol{v}}+(\boldsymbol{\nabla}\hat{\boldsymbol{v}})^{\mathrm{T}})$.
To exploit the reciprocal theorem, we first expand the velocity, pressure, and
stress fields into perturbation series in the dimensionless parameter
$\beta=u_{c}/h_{0}\ll1$ controlling the compliance of the
top wall:
\begin{subequations}\begin{align}
\boldsymbol{v}&=\hat{\boldsymbol{v}}+\beta\boldsymbol{v}_{1}+\mathrm{O}(\beta^{2}),\label{Velocity}\\p &= \hat{p} + \beta p_1 + \mathrm{O}(\beta^2),\label{Pressure}\\\boldsymbol{\sigma}_{f}&=\hat{\boldsymbol{\sigma}}_{f}+\beta\boldsymbol{\sigma}_{f,1}+\mathrm{O}(\beta^{2}),\label{Stress}
\end{align}\label{Expansion}\end{subequations}
where $\boldsymbol{v}_{1}$, $p_{1}$, and $\boldsymbol{\sigma}_{f,1}$
are the first-order corrections to the velocity, pressure, and hydrodynamic
stress in the \textit{rectangular domain} due to the deformation of
the top wall. From (\ref{Continuity+Momentum}) and (\ref{Expansion}),
it follows that the corresponding governing equations are the Stokes
equations $\boldsymbol{\nabla\cdot}\boldsymbol{v}_{1}=0$, $\boldsymbol{\nabla\cdot}\boldsymbol{\sigma}_{f,1}=\boldsymbol{0}$.
The Lorentz reciprocal theorem states that the two sets of velocity and stress fields $(\hat{\boldsymbol{v}},\hat{\boldsymbol{\sigma}}_{f})$
and $(\boldsymbol{v}_{1},\boldsymbol{\sigma}_{f,1})$ satisfy \cite{happel1983}:
\begin{equation}
\int_{S_{0}}\boldsymbol{n}\boldsymbol{\cdot}\boldsymbol{\sigma}_{f,1}\boldsymbol{\cdot}\hat{\boldsymbol{v}}\,\mathrm{d}S+\int_{S_{\ell}}\boldsymbol{n}\boldsymbol{\cdot}\boldsymbol{\sigma}_{f,1}\boldsymbol{\cdot}\hat{\boldsymbol{v}}\,\mathrm{d}S+\underbrace{\int_{S_{\mathrm{top}}}\boldsymbol{n}\boldsymbol{\cdot}\boldsymbol{\sigma}_{f,1}\boldsymbol{\cdot}\hat{\boldsymbol{v}}\,\mathrm{d}S}_{=0}=\int_{S_{0}}\boldsymbol{n}\boldsymbol{\cdot}\hat{\boldsymbol{\sigma}}_{f}\boldsymbol{\cdot}\boldsymbol{v}_{1}\,\mathrm{d}S+\int_{S_{\ell}}\boldsymbol{n}\boldsymbol{\cdot}\hat{\boldsymbol{\sigma}}_{f}\boldsymbol{\cdot}\boldsymbol{v}_{1}\,\mathrm{d}S+\int_{S_{\mathrm{top}}}\boldsymbol{n}\boldsymbol{\cdot}\hat{\boldsymbol{\sigma}}_{f}\boldsymbol{\cdot}\boldsymbol{v}_{1}\,\mathrm{d}S,\label{RT for deformable2}
\end{equation}
where $S_{\mathrm{top}}$ is the lower surface of the top wall \textit{in the undeformed state}, $S_{0}$ and $S_{\ell}$ are
the surfaces at the inlet ($z=0$) and outlet ($z=\ell$), respectively,
and $\boldsymbol{n}$ is the unit outward normal on a respective surface. Note that the integrals over the bottom and side walls of the channel vanish because  $\hat{\boldsymbol{v}}=\boldsymbol{v}_{1}=\boldsymbol{0}$ there. Also, the last term on the left-hand side of (\ref{RT for deformable2}) vanishes because $\hat{\boldsymbol{v}}=\boldsymbol{0}$ on $S_{\mathrm{top}}$.

Using the asymptotic expansion (\ref{Expansion}) and non-dimensionalization (\ref{Non-dimensional variables}),
we obtain 
\begin{equation}
\left.\boldsymbol{n}\boldsymbol{\cdot}\boldsymbol{\sigma}_{f,1}\boldsymbol{\cdot}\hat{\boldsymbol{v}}\right|_{z=0,\,\ell}=\mp\frac{\mu v_{c}^{2}\ell}{h_{0}^{2}}\left[-P_{1}\hat{V}_{z}+\mathrm{O}(\beta)\right]_{Z=0,\,1},\qquad\left.\boldsymbol{n}\boldsymbol{\cdot}\hat{\boldsymbol{\sigma}}_{f}\boldsymbol{\cdot}\boldsymbol{v}_{1}\right|_{z=0,\,\ell}=\mp\frac{\mu v_{c}^{2}\ell}{h_{0}^{2}}\left[-\hat{P}V_{z,1}+\mathrm{O}(\beta)\right]_{Z=0,\,1},\label{Scaling 1&2}
\end{equation}
where the minus sign in (\ref{Scaling 1&2}) corresponds
to $S_{0}$ and the plus sign corresponds to $S_{\ell}$ (see \cite{boyko2021RT,boyko2022pressure}). 

Similarly,
the integrand in the last term of (\ref{RT for deformable2}),
evaluated at $y=h_{0}$ (i.e., on $S_{\mathrm{top}}$), with $\boldsymbol{n}=\boldsymbol{e}_{y}$
and $\boldsymbol{n}\boldsymbol{\cdot}\hat{\boldsymbol{\sigma}}_{f}=\mu(\partial\hat{v}_{z}/\partial y)\boldsymbol{e}_{z}-\hat{p}\boldsymbol{e}_{y}$, is 
\begin{equation}
\left.\boldsymbol{n}\boldsymbol{\cdot}\hat{\boldsymbol{\sigma}}_f\boldsymbol{\cdot}\boldsymbol{v}_{1}\right|_{y=h_{0}}=\frac{\mu v_{c}^{2}}{h_{0}}\left[\frac{\partial\hat{V}_{z}}{\partial Y}V_{z,1}\right]_{Y=1}.\label{Scaling 3}
\end{equation}
Note that the pressure term in $\boldsymbol{n}\boldsymbol{\cdot}\hat{\boldsymbol{\sigma}}_{f}$ does not contribute to (\ref{Scaling 3}) because the leading-order flow is purely axial, so that $\left.\boldsymbol{v}_{1}\boldsymbol{\cdot}\boldsymbol{e}_{y}\right|_{y=h_{0}}=0$ due to no penetration.
We determine $V_{z,1}(X,Y=1,Z)$ by applying the no-slip boundary
condition, $V_{z}(X,Y=H(X,Z),Z)=0$, and using (\ref{Non-dimensional shape H})
and (\ref{Velocity}), together with the domain perturbation expansion
in $\beta$ introduced above, to obtain
\begin{equation}
V_{z}(X,H(X,Z),Z)=\hat{V}_{z}(X,1,Z)+\beta\left(V_{z,1}(X,1,Z)+U(X,Z)\left.\frac{\partial\hat{V}_{z}}{\partial Y}\right|_{Y=1}\right)+\mathrm{O}(\beta^2)=0.
\label{V_z expansion}
\end{equation}
It follows that $V_{z,1}(X,1,Z)=-U(X,Z)\left.\partial\hat{V}_{z}/\partial Y\right|_{Y=1}$. Thus, (\ref{Scaling 3}) reduces to 
\begin{equation}
\left.\boldsymbol{n}\boldsymbol{\cdot}\hat{\boldsymbol{\sigma}}_{f}\boldsymbol{\cdot}\boldsymbol{v}_{1}\right|_{y=h_{0}}=\frac{\mu v_{c}^{2}}{h_{0}}\left[\frac{\partial\hat{V}_{z}}{\partial Y}V_{z,1}\right]_{Y=1}=-\frac{\mu v_{c}^{2}}{h_{0}}U(X,Z)\left(\left.\frac{\partial\hat{V}_{z}}{\partial Y}\right|_{Y=1}\right)^{2}+\mathrm{O}(\beta).\label{Scaling 3 explicit}
\end{equation}
Substituting (\ref{Scaling 1&2}) and (\ref{Scaling 3 explicit}) into
(\ref{RT for deformable2}), and using the outlet boundary condition  $P_{1}(Z=1)=\hat{P}(Z=1)=0$, we obtain
\begin{equation}
\int_{0}^{1}\int_{-1/2}^{1/2}\left[P_{1}\hat{V}_{z}-\hat{P}V_{z,1}\right]_{Z=0}\mathrm{d}X\mathrm{d}Y=-\int_{0}^{1}\int_{-1/2}^{1/2}\left[U(X,Z)\left(\left.\frac{\partial\hat{V}_{z}}{\partial Y}\right|_{Y=1}\right)^{2}\right]\mathrm{d}X\mathrm{d}Z+\mathrm{O}(\beta).\label{RT ND}
\end{equation}
Noting that $P_{1}=P_{1}(Z)$ and $\hat{P}=\hat{P}(Z)$ to the leading order in $\epsilon$, consistent with the classical lubrication approximation \cite{leal2007advanced},
and $\int_{0}^{1}\int_{-1/2}^{1/2}\hat{V}_{z}\mathrm{d}X\mathrm{d}Y=1$
and $\int_{0}^{1}\int_{-1/2}^{1/2}V_{z,1}\mathrm{d}X\mathrm{d}Y=0$, (\ref{RT ND})
yields the first-order correction to the pressure drop, defined as $\Delta P_{1}=P_{1}(Z=0)$,
for the weakly deformable channel ($\beta\ll1$):
\begin{equation}
\Delta P_{1}=-\int_{0}^{1}\int_{-1/2}^{1/2}\left[U(X,Z)\left(\left.\frac{\partial\hat{V}_{z}}{\partial Y}\right|_{Y=1}\right)^{2}\right]\mathrm{d}X\mathrm{d}Z.\label{dP1-Q relation}
\end{equation}
Equation (\ref{dP1-Q relation}) is the first key result of this Letter, which allows the determination of the first-order correction to the pressure drop of the deformable channel, provided the top wall deformation $U(X,Z)$ is known. Therefore, in fact, (\ref{dP1-Q relation}) is not restricted to deformable channels, and provides the first-order correction to the pressure drop for the three-dimensional rigid channel, whose top wall is non-uniform and has any prescribed shape variation expressed as $H(X,Z)=1+\beta U(X,Z)$.

\subsection{Elastic reciprocal theorem}

Let $\tilde{\boldsymbol{u}}$ and $\tilde{\boldsymbol{\sigma}}_{s}$ denote, respectively, the displacement and stress fields corresponding to the solution of the elastic problem in the same domain but with different boundary conditions on the stress or displacement fields. The corresponding governing equation is $\boldsymbol{\nabla\cdot}\tilde{\boldsymbol{\sigma}}_{s}=\boldsymbol{0}$, with $\tilde{\boldsymbol{\sigma}}_{s}=\lambda(\boldsymbol{\nabla\cdot}\tilde{\boldsymbol{u}})\boldsymbol{I}+2G\tilde{\boldsymbol{D}}$. The  Maxwell--Betti reciprocal theorem \cite{betti1872teoria,maxwell1864calculation,achenbach2003reciprocity} states that the solutions,  $(\boldsymbol{u},\boldsymbol{\sigma}_{s})$ and $(\tilde{\boldsymbol{u}},\tilde{\boldsymbol{\sigma}}_{s})$, to the two elastostatic problems satisfy:
\begin{equation}
\int_{S}\boldsymbol{n}\boldsymbol{\cdot}\boldsymbol{\sigma}_{s}\boldsymbol{\cdot}\tilde{\boldsymbol{u}}\,\mathrm{d}S=\int_{S}\boldsymbol{n}\boldsymbol{\cdot}\tilde{\boldsymbol{\sigma}}_{s}\boldsymbol{\cdot}\boldsymbol{u}\,\mathrm{d}S,
\label{Reciprocal theorem elastic}
\end{equation}
where $\boldsymbol{n}$ is the unit outward normal to the bounding surfaces $S$ of the elastic solid.

Before applying the elastic reciprocal theorem (\ref{Reciprocal theorem elastic}) to our elastohydrodynamic problem, recall a few assumptions we have made. First, in this problem, $\boldsymbol{u}=u_{y}(x,z)\boldsymbol{e}_{y}$ and $\tilde{\boldsymbol{u}}=\tilde{u}_{y}(x,z)\boldsymbol{e}_{y}$. Second, there is no-displacement on the lateral walls of the solid. Third, the continuity of tractions, $\boldsymbol{n}\boldsymbol{\cdot}\boldsymbol{\sigma}_{s}=\boldsymbol{n}\boldsymbol{\cdot}\boldsymbol{\sigma}_{f}$, holds at the fluid$-$solid interface, $y=h(x,z)$. Lastly, we have assumed a stress-free condition at the upper surface of the top
wall. Based on these
assumptions, the terms, $\boldsymbol{n}\boldsymbol{\cdot}\boldsymbol{\sigma}_{s}\boldsymbol{\cdot}\tilde{\boldsymbol{u}}$ and $\boldsymbol{n}\boldsymbol{\cdot}\tilde{\boldsymbol{\sigma}}_{s}\boldsymbol{\cdot}\boldsymbol{u}$, appearing in (\ref{Reciprocal theorem elastic}) are calculated as
\begin{subequations}
\begin{gather}
\left.\boldsymbol{n}\boldsymbol{\cdot}\boldsymbol{\sigma}_{s}\boldsymbol{\cdot}\tilde{\boldsymbol{u}}\right|_{y=h(x,z)}=\left.\boldsymbol{n}\boldsymbol{\cdot}\boldsymbol{\sigma}_{f}\boldsymbol{\cdot}\tilde{\boldsymbol{u}}\right|_{y=h(x,z)}
=
p_{c}u_{c}\left[-P\tilde{U}+\mathrm{O}(\epsilon)\right]=p_{c}u_{c}\left[-\hat{P}\tilde{U}+\mathrm{O}(\epsilon,\beta)\right],\label{Scaling Elasticty 1}\\
\left.\boldsymbol{n}\boldsymbol{\cdot}\tilde{\boldsymbol{\sigma}}_{s}\boldsymbol{\cdot}\boldsymbol{u}\right|_{y=h(x,z)}=\left.\boldsymbol{n}\boldsymbol{\cdot}\tilde{\boldsymbol{\sigma}}_{f}\boldsymbol{\cdot}\boldsymbol{u}\right|_{y=h(x,z)}
=
p_{c}u_{c}\left[-\tilde{P}U+\mathrm{O}(\epsilon)\right]=p_{c}u_{c}\left[-\hat{\tilde{P}}U+\mathrm{O}(\epsilon,\beta)\right],
\label{Scaling Elasticty 2}
\end{gather}\label{Scaling Elasticty}\end{subequations}
where to obtain the last equality we used the domain perturbation expansion introduced above, i.e., $P = \hat{P} + \mathrm{O}(\beta)$ and $\tilde{P} = \hat{\tilde{P}} + \mathrm{O}(\beta)$.
Substituting (\ref{Scaling Elasticty}) into (\ref{Reciprocal theorem elastic}) leads to
\begin{equation}
\int_{0}^{1}\int_{-1/2}^{1/2}\hat{\tilde{P}}U\,\mathrm{d}X\mathrm{d}Z=\int_{0}^{1}\int_{-1/2}^{1/2}\hat{P}\tilde{U}\,\mathrm{d}X\mathrm{d}Z.\label{Reciprocal theorem elastic FSI-2}
\end{equation}
Next, we utilize the convolution principle to obtain an explicit expression for the deformation $U(X,Z)$ from (\ref{Reciprocal theorem elastic FSI-2}). Choosing $\hat{\tilde{P}}$ as a point load, i.e., $\hat{\tilde{P}}=\delta_{D}(X-\mathcal{X})\delta_{D}(Z-\mathcal{Z})$, applied on the fluid--solid interface at the point $(\mathcal{X},\mathcal{Z})$, where $\delta_{D}$ is the Dirac delta distribution, we obtain
\begin{equation}
U(X,Z)=\int_{0}^{1}\int_{-1/2}^{1/2}\hat{P}(\mathcal{X},\mathcal{Z})\tilde{U}(X,Z;\mathcal{X},\mathcal{Z})\,\mathrm{d}\mathcal{X}\mathrm{d}\mathcal{Z}.
\label{Reciprocal theorem elastic FSI-3}
\end{equation}
Here, $\tilde{U}(X,Z;\mathcal{X},\mathcal{Z})$ is the point-load
solution (or, Green's function) of the elastic problem with appropriate
boundary conditions under the aforementioned assumptions.

\subsection{Flow rate--pressure drop relation for deformable channels using fluidic and elastic reciprocal theorems}

Combining (\ref{dP1-Q relation}) and (\ref{Reciprocal theorem elastic FSI-3}), we obtain the first-order correction to the pressure drop for the
weakly deformable channel, $\beta\ll1$, expressed using the fluidic
and elastic reciprocal theorems:
\begin{equation}
\Delta P_{1}=-\int_{0}^{1}\int_{-1/2}^{1/2}\left[\int_{0}^{1}\int_{-1/2}^{1/2}\hat{P}(\mathcal{X},\mathcal{Z})\tilde{U}(X,Z;\mathcal{X},\mathcal{Z}) \,\mathrm{d}\mathcal{X}\mathrm{d}\mathcal{Z}\left(\left.\frac{\partial\hat{V}_{z}}{\partial Y}\right|_{Y=1}\right)^{2}\right]\mathrm{d}X\mathrm{d}Z.
\label{dP1-Q relation general}
\end{equation}
Equation (\ref{dP1-Q relation general}) is the second key result of this Letter, clearly indicating that the first-order correction to the pressure drop arises due to the interplay between the fluid stresses and the compliance of the channel. Furthermore, (\ref{dP1-Q relation general}) shows that $\Delta P_{1}$ depends only on the solution of the pressure-driven flow in a rigid channel for the fluid problem (hat) and the solution of the elastic deformation for a point-load  (tilde), thus eliminating the need to \textit{a priori} solve the coupled elastohydrodynamic problem.

The dimensionless pressure drop is thus $\Delta P=\Delta\hat{P}+\beta\Delta P_{1}+\mathrm{O}(\beta^{2})$, where the solution of the corresponding pressure-driven flow in a rigid (rectangular) channel \cite{bruus2008book} is well known:
\begin{subequations}
\begin{gather}
\hat{P}(Z)=\frac{12}{1-\kappa(\delta)}(1-Z),\qquad\Delta\hat{P}=\frac{12}{1-\kappa(\delta)},\qquad\text{where}\;\;\kappa(\delta)=\frac{192}{\pi^{5}}\sum_{n=1}^{\infty}\frac{1}{(2n-1)^{5}}\delta\tanh\left[\frac{(2n-1)\pi}{2\delta}\right],\label{P(Z)_hat}\\
\hat{V}_{z}(X,Y)=\frac{12}{1-\kappa(\delta)}\left[\frac{1}{2}(1-Y)Y-\frac{4}{\pi^{3}}\sum_{n=1}^{\infty}\frac{1}{(2n-1)^{3}}\frac{\cosh[(2n-1)\pi X/\delta]}{\cosh[(2n-1)\pi/(2\delta)]}\sin[(2n-1)\pi Y]\right].\label{Vz(X,Y)_hat}
\end{gather}\label{Hat problem}\end{subequations}

\section{Illustrated example}

\begin{figure}[t]
 \centerline{\includegraphics[scale=1.2]{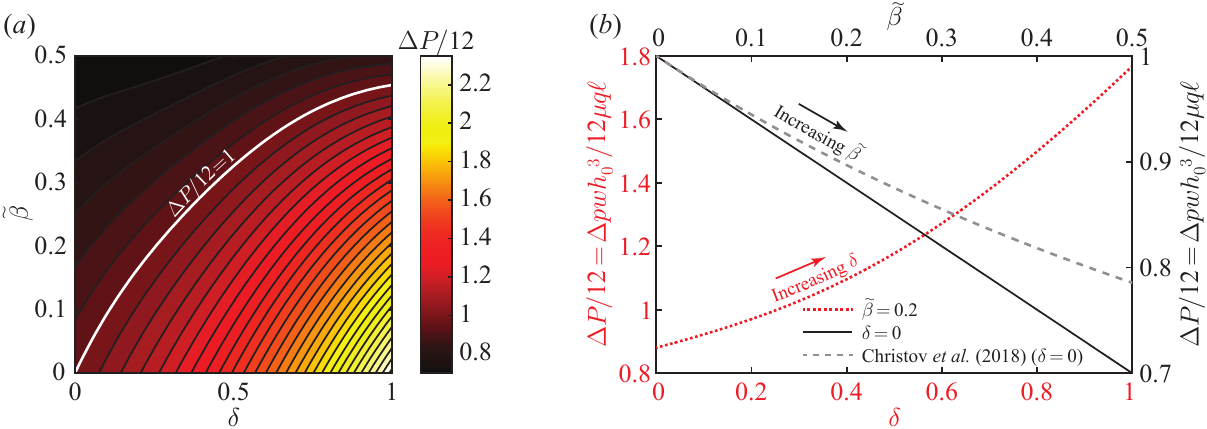}}\caption{Dimensionless pressure drop $\Delta P/12=\Delta pwh_{0}^{3}/12\mu q\ell$ in a deformable channel modeled using the plate-bending theory. (a) Contour plot of the dimensionless pressure drop as a function of $\tilde{\beta}=u_{c}/24h_{0}$ and $\delta=h_{0}/w$. (b) Dimensionless pressure drop as a function of $\delta=h_{0}/w$ (dotted line) for $\tilde{\beta}=0.2$ and as a function of $\tilde{\beta}=u_{c}/24h_{0}$ (solid and dashed curves) for $\delta=0$. The solid line represents the first-order asymptotic solution and the dashed line represents the solution based on the lubrication theory (also for $\delta=0$), derived by \citet{christov2018flow}.}
\label{F2}
\end{figure}

In this section, we illustrate an application of our results towards calculating the first-order-in-$\beta$ correction to the pressure drop in a slender compliant channel. For this example and for simplicity, we model the compliance of the top wall using the plate-bending theory. Under the assumptions that the maximum displacement of the top wall is small compared to its thickness $b$, and the thickness $b$ is small compared to its width $w$, $\max_{x,z}[u_{y}(x,z)]\ll b\ll w$, the steady-state displacement $u_{y}(x,y)$ satisfies the Kirchhoff--Love equation for isotropic bending of a plate under a transverse load supplied by the fluid pressure, i.e., $p=B\nabla_{\Vert}^{4}u_{y}$ \cite{love1888xvi,timoshenkoPlates,howell2009applied}. Here, $B=Eb^{3}/[12(1-\nu^{2})]$ is the bending stiffness, and $\nabla_{\Vert}^{4}$ is the biharmonic operator in the $(x,z)$ plane. 

Using (\ref{Non-dimensional variables}) and performing order-of-magnitude analysis, we obtain $u_{c}=w^{4}p_{c}/B=w^{3}\mu q\ell/(h_{0}^{3}B)$ and thus $\beta=u_{c}/h_{0}=w^{4}p_{c}/(Bh_{0})=w^{3}\mu q\ell/(h_{0}^{4}B)$. Furthermore, it follows that $P(Z)=\partial^{4}U/\partial X^{4}+\mathrm{O}(\epsilon^{2}/\delta^{2})$
\cite{christov2018flow,martinez2020start}, where $0\ll\epsilon\ll\mathrm{min}(1,\delta)$ but $\delta=h_{0}/w$ is not necessary small. The Green's function corresponding to the point-load solution of the latter equation with clamped boundary conditions \cite{duffy2015green}, i.e., $U|_{X=\pm1/2}=\left.\partial U/\partial X\right|_{X=\pm1/2}=0$, is 
\begin{equation}
\tilde{U}(X,Z;\mathcal{X},\mathcal{Z})=\frac{1}{24}\left(X-\frac{1}{2}\right)^{2}\left(X+\frac{1}{2}\right)^{2}\delta_{D}(Z-\mathcal{Z}).\label{Green's function}
\end{equation}
Using (\ref{Hat problem}) and (\ref{Green's function}), (\ref{dP1-Q relation general}) provides the first-order correction to the pressure drop for arbitrary value of $\delta=h_{0}/w$. We note that it is difficult to obtain a closed-form expression for $\Delta P_{1}$, which holds for any $\delta=h_{0}/w$, because of the velocity profile (\ref{Vz(X,Y)_hat}) is represented as an infinite series. While the $Z$-integration can be performed analytically, the $X$-integration is done numerically. However, for $\delta\ll1$, from (\ref{Vz(X,Y)_hat}) it follows that $\Delta\hat{P}=12$, $\hat{P}(Z)=12(1-Z)$, and $\hat{V}_{z}(Y)=6(1-Y)Y$, and substituting theses results and (\ref{Green's function}) into (\ref{dP1-Q relation general}) yields $\Delta P_{1}/12=-3/120$, so that $\Delta P /12=1-(3/120)\beta+\mathrm{O}(\beta^{2})=1-(3/5)\tilde{\beta}+\mathrm{O}(\tilde{\beta}^{2})$, where $\tilde{\beta}=\beta/24=w^{4}p_{c}/(24Bh_{0})=w^{3}\mu q\ell/(24h_{0}^{4}B)$. 

It is instructive to compare the latter result to the solution for the dimensionless pressure drop previously derived by \citet{christov2018flow} using lubrication theory. The result from \cite{christov2018flow}, holding for $\delta\ll1$ and $\tilde{\beta}=\mathrm{O}(1)$, is expressed as an implicit relation:
\begin{equation}
12=\Delta P\left[1+\frac{\tilde{\beta}}{20}\Delta P+\frac{\tilde{\beta}^{2}}{630}(\Delta P)^{2}+\frac{\tilde{\beta}^{3}}{48048}(\Delta P)^{3}\right]\qquad(\mathrm{lubrication\,theory\,for}\,\delta\ll1),
\label{Pressure drop Christov}
\end{equation}
and it also yields $\Delta P/12=1-(3/5)\tilde{\beta}+(86/175)\tilde{\beta}^{2}+\mathrm{O}(\tilde{\beta}^{3})$ for $\beta\ll1$, upon solving for the positive real root of \eqref{Pressure drop Christov}. Observe that this expression is identical to our asymptotic solution for the pressure drop to first order in $\tilde{\beta}$.

In Fig.~\ref{F2}(a), we present a contour plot of the dimensionless pressure drop  accounting for the first-order correction due to fluid--structure interaction as a function of $\tilde{\beta}=u_{c}/24h_{0}$ and $\delta=h_{0}/w$, for the deformable channel modeled using the plate-bending theory. Figure~\ref{F2}(a) clearly indicates the existence of the trade-off between the effect of compliance of the deforming top wall and the effect of sidewalls on the pressure drop. While the pressure drop decreases with $\tilde{\beta}$, it increases with $\delta$. The white (light) solid curve $\Delta P/12=1$ divides the colormap into two regions: in the upper region, the compliance dominates over the sidewall effects and thus $\Delta P/12<1$, whereas in the lower region, the sidewall drag is dominant and $\Delta P /12>1$.

Next, in Fig.~\ref{F2}(b), we show a comparison of our analytical predictions and the lubrication-theory-based solution (\ref{Pressure drop Christov}) for the non-dimensional pressure drop as a function of $\tilde{\beta}$, for $\delta=0$ (an infinitely wide channel). The black solid line represents our first-order asymptotic solution, $\Delta P/12=1-(3/5)\tilde{\beta}+\mathrm{O}(\tilde{\beta}^{2})$, and the gray dashed curve represents the lubrication solution, (\ref{Pressure drop Christov}). It is evident from Fig.~\ref{F2}(b) that our first-order asymptotic solution, which is strictly valid for $\tilde{\beta}\ll1$, slightly underpredicts the lubrication solution; yet even for $\tilde{\beta}=0.5$, it results in a modest relative error of approximately 10$\%$. For further clarification, the dotted (red) curve shows the dimensionless pressure drop as a function of $\delta=h_{0}/w$, for $\tilde{\beta}=0.2$, clearly indicating that, while increased wall compliance decreases the pressure drop, drag due to the sidewalls increases it.

\section{Concluding remarks}

In this Letter, we showed how the reciprocal theorems for Stokes flow and linear elasticity can be used to derive a closed-form expression for leading-order correction, due to deformation, to the flow rate--pressure drop relation for rectangular channels. Using a domain perturbation approach and considering small deformation, our theory captures the interplay between the fluid stresses and the compliance of the channel's top wall, bypassing the need to calculate the solution of the coupled fluid-structure-interaction problem. Furthermore, unlike previous theoretical studies, which neglected the presence of lateral sidewalls (and considered shallow geometries of effectively infinite width within the lubrication approximation), our approach allows the determination of the influence of confining sidewalls on the $q-\Delta p$ relation.

The present theoretical approach is not limited to the case of a three-dimensional channel of initially rectangular cross-section, and it could also be applied to calculate the first-order correction to the pressure drop in axisymmetric deformable tubes, for which the leading-order pressure drop is given by the classical Hagen--Poiseuille law \cite{sutera1993history}. Finally, while we considered viscous Newtonian fluids, it would be of interest to understand how the rheological response of complex fluids (such as shear thinning and viscoelasticity) influences the flow rate--pressure drop relation in deformable channels. One convenient approach to accomplish this task would be to rely on reciprocal theorems, and use a combination of the present approach and the approach recently established by~\citet{boyko2021RT,boyko2022pressure} for calculating the effect of complex fluid rheology on the flow rate--pressure drop relation for rigid non-uniform channels. These calculations are left for future investigation.

\begin{acknowledgments}
E.B.\ acknowledges the support of the Yad Hanadiv (Rothschild) Foundation, the Zuckerman STEM Leadership Program, and the Lillian Gilbreth Postdoctoral Fellowship from Purdue’s College of Engineering. H.A.S.\ is grateful for partial support of the work via NSF grant CBET-2127563. I.C.C.\ acknowledges partial support by the NSF under grant No.\ CBET-1705637 in the early stages of this work.
\end{acknowledgments}

%


\end{document}